\documentclass[useAMS,usenatbib]{mn2e}
\usepackage{graphicx,amssym,color}
\newcommand{\Mpch}{\mbox{ $h^{-1}$ Mpc}}

\newcommand{\be}{\begin{equation}}
\newcommand{\ee}{\end{equation}}

\def\ltsima{$\; \buildrel < \over \sim \;$}
\def\simlt{\lower.5ex\hbox{\ltsima}}
\def\gtsima{$\; \buildrel > \over \sim \;$}
\def\simgt{\lower.5ex\hbox{\gtsima}}
\def\la{$\langle$}

\title{The Mass Profile and Accretion History of Cold Dark Matter Halos}

\author[Ludlow et al.] {\parbox{18cm}{
Aaron D. Ludlow$^{1,\star}$,
Julio F. Navarro$^{2}$,
Michael Boylan-Kolchin$^{3}$,
Philip E. Bett$^{1}$,
Ra\'ul E. Angulo$^{4}$,
Ming Li$^{5,6}$,
Simon D. M. White$^{5}$,
Carlos Frenk$^{7}$,
Volker~Springel$^{8,9}$
}\vspace{0.3cm}\\
$^{1}${Argelander-Institut f\"{u}r Astronomie, Auf dem H\"{u}gel 71,
  D-53121 Bonn, Germany}\\
$^{2}${Dept. of Physics and Astronomy, University of
  Victoria, Victoria, BC, V8P 5C2, Canada}\\
$^{3}${Center for Galaxy Evolution, 4129 Reines Hall, University of  
  California, Irvine, CA 92697, USA}\\
$^{4}${Kavli Institute for Particle Astrophysics and Cosmology, Stanford University,
  SLAC National Laboratory, Menlo Park, CA 94025, USA}\\
$^{5}${Max-Planck-Institut f\"{u}r Astrophysik,
  Karl-Schwarzschild-Stra\ss{}e 1, 85740 Garching bei M\"{u}nchen, Germany}\\
$^{6}${Purple Mountain Observatory, West Beijing Rd. 2, 
  210008 Nanjing, China} \\
$^{7}${Institute for Computational Cosmology, Dept. of Physics, Univ. of
  Durham, South Road, Durham  DH1 3LE, UK}\\
$^{8}${Heidelberg Institute for Theoretical Studies,
  Schloss-Wolfsbrunnenweg 35, 69118 Heidelberg, Germany}\\
$^{9}${Zentrum f\"{u}r Astronomie der Universit\"{a}t Heidelberg, ARI,
  M\"{o}nchhofstr. 12-14, 69120 Heidelberg, Germany}\\
}

\begin{document}

\maketitle 

\begin{abstract}
  We use the Millennium Simulation series to investigate the relation
  between the accretion history and mass profile of cold dark matter
  halos. We find that the mean inner density within the scale radius,
  $r_{-2}$ (where the halo density profile has isothermal slope), is
  directly proportional to the critical density of the Universe at the
  time when the virial mass of the main progenitor equals the mass
  enclosed within $r_{-2}$. Scaled to these characteristic values of
  mass and density, the average mass accretion history, expressed in
  terms of the critical density of the Universe, $M(\rho_{\rm
    crit}(z))$, resembles that of the enclosed density profile,
  $M(\langle \rho \rangle)$, at $z=0$. Both follow closely the NFW
  profile, which suggests that the similarity of halo mass profiles
  originates from the mass-independence of halo accretion
  histories. Support for this interpretation is provided by outlier
  halos whose accretion histories deviate from the NFW shape; their
  mass profiles show correlated deviations from NFW and are better
  approximated by Einasto profiles. Fitting both $M(\langle \rho
  \rangle)$ and $M(\rho_{\rm crit})$ with either NFW or Einasto
  profiles yield concentration and shape parameters that are
  correlated, confirming and extending earlier work that has linked
  the concentration of a halo with its accretion history. These
  correlations also confirm that halo structure is insensitive to
  initial conditions: only halos whose accretion histories differ
  greatly from the NFW shape show noticeable deviations from NFW in
  their mass profiles.  As a result, the NFW profile provides
  acceptable fits to hot dark matter halos, which do not form
  hierarchically, and for fluctuation power spectra other than
  CDM. Our findings, however, predict a subtle but systematic
  dependence of mass profile shape on accretion history which, if
  confirmed, would provide strong support for the link between
  accretion history and halo structure we propose here.
\end{abstract}

\renewcommand{\thefootnote}{\fnsymbol{footnote}}
\footnotetext[1]{E-mail: aludlow@astro.uni-bonn.de}

\section{Introduction}
\label{SecIntro}

Numerical simulations have shown that the equilibrium structure of
cold dark matter (CDM) halos is approximately
self-similar. Spherically averaged density profiles, in particular,
are well approximated by scaling a simple formula proposed by
\citet[hereafter NFW]{Navarro1995,Navarro1996}. The NFW profile has
fixed shape, and is characterized by a logarithmic slope that steepens
gradually from the center outwards. As such, it may be fully specified
by just two parameters, usually chosen to be either the virial radius
and a characteristic density or, equivalently, the halo virial mass
and a concentration parameter. (See Sec.~\ref{SecFitForm} for a
summary of relevant formulae and definitions.)

The gently-varying slope of the NFW profile confounded early
theoretical expectations, which had envisioned a simple power-law
behaviour
\citep[e.g.,][]{Fillmore1984,Hoffman1985,Quinn1986,Crone1994}, and has
motivated a number of proposals to explain its origin \citep[see, for
a recent review,][]{Frenk2012}. Most attempts have taken as guidance
the secondary-infall model first proposed by \citet{Gunn1972},
complemented by various conjectures about the role of mergers
\citep[e.g.,][]{Salvador-Sole1998}, dynamical friction
\citep[e.g.,][]{Nusser1999}, angular momentum
\citep[e.g.,][]{Williams2004}, or adiabatic invariants
\citep[e.g.,][]{Avila-Reese1998,Dalal2010}, or else have argued that
entropy generation during virialization might be behind the halo
structural similarity \citep[see, e.g.,][]{Taylor2001,Pontzen2013}.

No general consensus has yet emerged, however, reflecting the
difficulty that all of these models face when trying to explain why
the same NFW profile seems to fit the structure of halos formed
through hierarchical clustering regardless of power spectrum
\citep{Navarro1997}, as well as that of hot dark matter halos or of
systems formed through monolithic collapse \citep[e.g.,][]{Huss1999,Wang2009}.

In addition, none of these models provides a thorough explanation for
the redshift-dependent correlations between mass and concentration
seen in simulations, their scatter, or their dependence on cosmological
parameters and power spectra. Halo concentration, which depends only
weakly on mass, was originally linked to halo collapse time
\citep{Navarro1997}, but attempts to reproduce the simulation results
with simple prescriptions based on that proposal have met with limited
success \citep[][]{Bullock2001,Eke2001,Neto2007,Maccio2008,Gao2008}.

Better results have been obtained with empirical models that relate
concentration to halo mass accretion history and, in particular, to the
time when the main halo progenitor switches from a period of ``fast
growth'' to one of ``slow growth''
\citep{Wechsler2002,Zhao2003a,Lu2006}. The success of these
models is not, however, unqualified. \citet{Zhao2009}, for example,
argue that halo concentration is determined at the time when the main
progenitor first reaches $4\%$ of the final mass, but there seems to
be no natural justification for why concentration should be related to
this particular, rather arbitrary time of a halo's assembly history.

Further complicating matters, there is now convincing evidence that a
number of halos have density profiles that deviate slightly, but
significantly, from the NFW profile \citep{Navarro2004}.  Accounting
for these deviations requires the introduction of an additional {\it
  shape} parameter, thus breaking the structural similarity of CDM
halos. One parameterization that results in excellent fits is the
Einasto profile, where the logarithmic slope is a simple power law of
radius, $d\ln\rho/d\ln r\propto (r/r_{-2})^{\alpha}$: the shape
parameter, $\alpha$, is the exponent of the power law. This finding
has now been verified by additional work
\citep{Merritt2005,Merritt2006,Navarro2010,Gao2008,Hayashi2008,Stadel2009,Ludlow2011}
but there is no clear understanding of what breaks the similarity or
what determines the value of $\alpha$ for a particular halo. 

We explore these issues here using a large ensemble of halos selected
from the three Millennium Simulations, MS-I \citep{Springel2005a},
MS-II \citep{Boylan-Kolchin2009}, and MS-XXL \citep{Angulo2012},
collectively referred to hereafter as MS. These are amongst the
largest cosmological $N$-body simulations available, and provide us
with thousands of well-resolved halos spanning more than four decades
in mass. Merger trees are available for all these simulations, making
them an ideal dataset to explore the relation between accretion
history and mass profiles. In addition, the numerical homogeneity and
sheer size of the volumes surveyed by the MS allow us to combine large
numbers of halos with similar properties to smooth out statistical
fluctuations and idiosyncrasies of individual systems that might
obscure the general trends. Our analysis reveals a subtle but
well-defined relation between mass profile and accretion history that
offers valuable new clues to the origin of the structure of CDM halos.

The plan of this paper is as follows. We describe the simulations in
Sec.~\ref{SecNumSims} and the analysis procedure in
Sec.~\ref{SecAnal}. We present our main results in Sec.~\ref{SecRes}
and summarize our main conclusions in Sec.~\ref{SecConc}.

\begin{figure*}
\begin{center}
\resizebox{16cm}{!}{\includegraphics{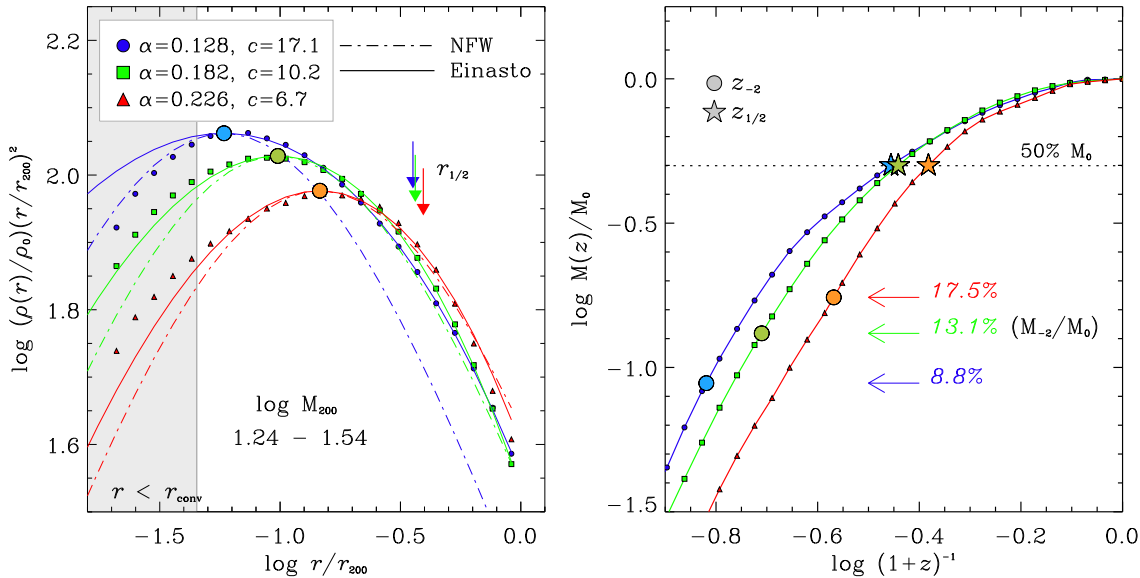}}
\end{center}
\caption{Halo density profiles and accretion histories. {\it Left
    panel:} Median density profiles of MS-II relaxed halos in the
  mass range $1.24 < \log M_{200}/(10^{10} h^{-1} M_{\odot}) < 1.54$
  (corresponding to particle numbers in the range $2.5\times 10^4 <
  N_{200} < 5\times 10^4$), selected according to their concentration
  (see boxes in the top panel of Fig.~\ref{FigMcA}). Densities are
  shown scaled to $\rho_0$, the critical density at $z=0$, and
  weighted by $r^2$ in order to enhance the dynamic range of the
  plot. Radii are scaled to the virial radius, $r_{200}$. The best-fit
  Einasto profiles are shown by the thin solid curves, with parameters
  listed in the legend. Dot-dashed curves indicate NFW profiles (whose
  shape is fixed in these units) matched at the scale radius, $r_{-2}$, where the
  $r^2\rho$ profiles peak. Arrows indicate the half-mass radius,
  $r_{1/2}$. {\it Right panel:} Median mass accretion histories (MAH) of
  the same set of halos chosen for the left panel. Halo accretion
  history is defined as the evolution of the mass of the main progenitor, expressed in
  units of the mass of the halo at $z=0$. The heavy circles indicate
  the redshift, $z_{-2}$, when the progenitor's mass equals the mass,
  $M_{-2}$, enclosed within the scale radius at $z=0$. Starred symbols
  indicate the half-mass formation redshift.}
\label{FigRhoProf}
\end{figure*}

\section{Numerical Simulations}
\label{SecNumSims}

Our analysis focuses on dark matter halos identified in the three
Millennium Simulations. We provide here a brief summary of these
simulations and of their associated halo catalogs. We refer the reader to
the original papers for extensive details on each of the MS runs.

\begin{figure}
\begin{center}
\resizebox{8cm}{!}{\includegraphics{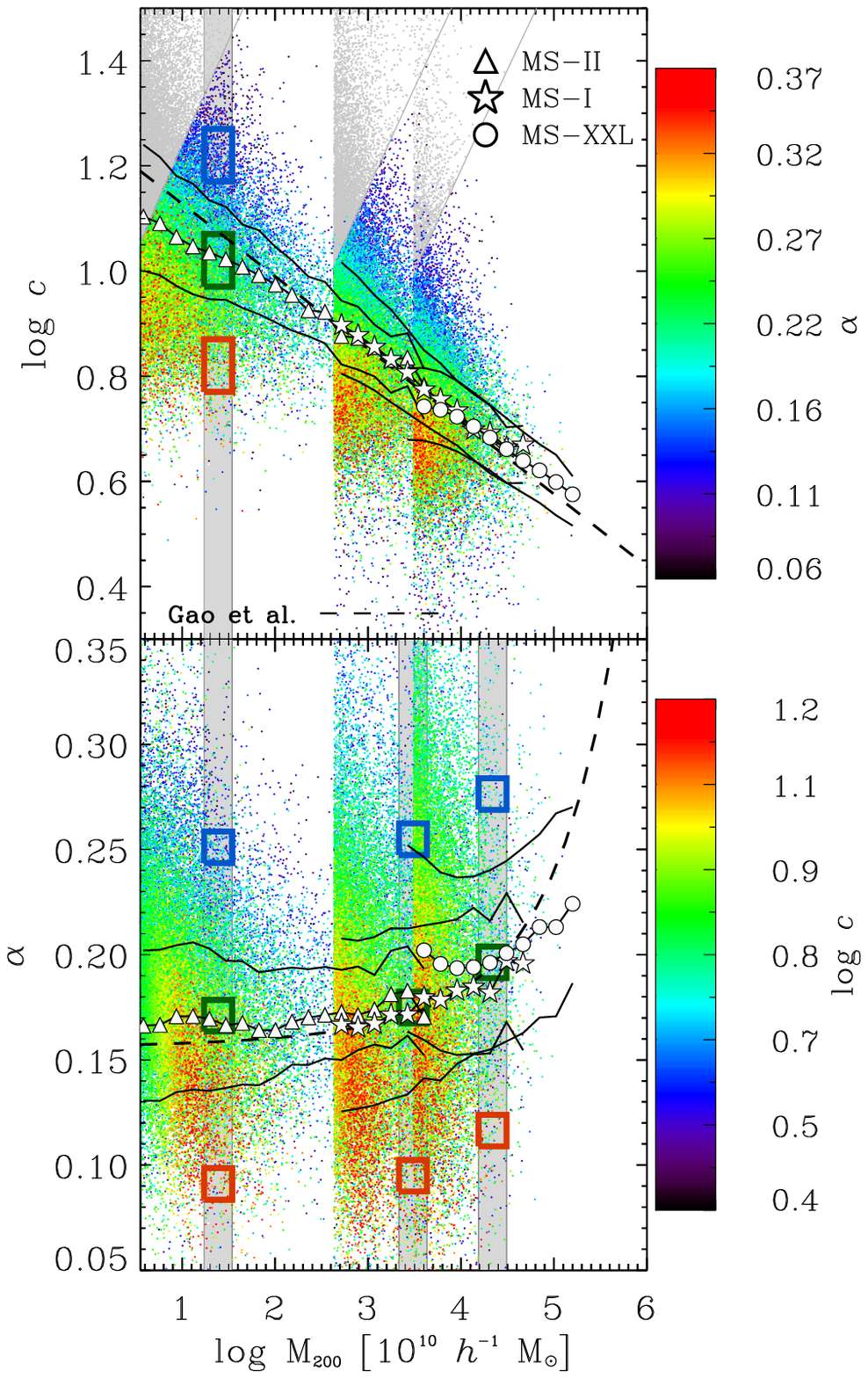}}
\end{center}
\caption{Mass dependence of the best-fit Einasto parameters for all
  halos in our sample at $z=0$. Only relaxed halos with more than $5000$
  particles within the virial radius are considered. The top and
  bottom panels show, respectively, the concentration,
  $c=r_{200}/r_{-2}$, and shape parameter, $\alpha$, as a function of
  halo virial mass. Individual points are colored according to the
  third parameter (see color bar on the right of each
  panel). Connected symbols trace the median values for each
  Millennium Simulation (see legend in the top panel); thin solid
  lines delineate the 25 to 75 percentile range. The dashed curves
  indicate the fitting formulae proposed by \citet{Gao2008}. For
  clarity only $10,000$ halos per simulation are shown in this figure.
  Halos shown in grey are systems where the best-fit scale radius is
  smaller than the convergence radius; these fits are deemed
  unreliable and the corresponding halos are not included in the
  analysis. Grey vertical bars highlight three different mass bins
  used to explore parameter variations at fixed halo mass (see
  Sec.~\ref{SecNFWMAH} and \ref{SecEMAH}). Small boxes indicate halos
  in each of those bins with average,
  higher-than-average, and lower-than-average values of $\alpha$
  (bottom panel) or of the concentration (top panel).}
\label{FigMcA}
\end{figure}

\subsection{The Millennium Simulation suite}
\label{SecMillSims}

All MS runs adopt a flat, WMAP 1-normalized LCDM cosmology with the
following cosmological parameters: $\Omega_{\rm m}=0.25$,
$\Omega_{\Lambda}=1-\Omega_{\rm m}=0.75$, $h=0.73$, $n=1$ and
$\sigma_8=0.9$. Here $\Omega_i$ is the present-day contribution of
component $i$ to the total matter energy density in units of the
critical density for closure, $\rho_{\rm crit}$; $\sigma_8$ is the rms mass
fluctuation in $8 \, \Mpch$ spheres, linearly extrapolated to $z=0$;
$n$ is the spectral index of primordial density fluctuations, and $h$
is the Hubble parameter. In addition to using the same cosmological
parameters, the MS runs also adopted the same sequence of outputs in
order to facilitate comparisons between the runs.
 
MS-II follows the dark matter distribution using 2160$^3$ particles of
mass $m_p=6.89\times10^6 \, h^{-1} M_{\odot}$ in a $100 \, h^{-1}{\rm
  Mpc}$ periodic box. MS-I has the same total particle number, but
follows the evolution of structure in a comoving box of $500 \,
h^{-1}$ Mpc on a side; each particle in MS-I is thus $125 \times$ more
massive than in MS-II, or $m_p=8.61\times10^8 \, h^{-1}
M_{\odot}$. MS-XXL is the largest of the three simulations in both box
size and particle number; it follows $6720^3$ particles of mass
$m_p=6.17\times10^9 \, h^{-1} M_{\odot}$ in a $3 \, h^{-1} {\rm Gpc}$
box.

\subsection{Halo Catalogs}
\label{SecHaloCats}

A friends-of-friends (FOF) group finder \citep[]{Davis1985} was run on
the fly for each simulation output using a linking length of $b=0.2$
times the mean inter-particle separation and a minimum particle number
$N_{\rm min}=20$. The subhalo finder \textsc{subfind},
\citep[]{Springel2001b} was then run to identify self-bound
substructure within each FOF halo. 

\textsc{Subfind} dissects each FOF halo into one dominant structure
(the main halo) and a number of subhalos that trace the self-bound
remnants of accreted systems.  We will focus our analysis on main
halos identified at $z=0$ that contain at least $N_{200}=5000$ particles
within their virial radius\footnote{We define the virial radius,
  $r_{200}$, of a halo as the radius of a sphere centered at the
  potential minimum that encloses a mean density of
  $200\times\rho_{\rm crit}$. We identify all virial quantities (i.e.,
  measured within $r_{200}$) with a ``200'' subscript. 
  Note that all particles are used to compute
  $r_{200}$, not just those bound to the main halo.}

Since dark matter halos are dynamical systems, transients induced by
mergers or ongoing accretion can lead to rapid fluctuations in the
structure of a halo that are poorly captured with simple fitting
formulae. We therefore impose three criteria to flag systems that are
clearly out of equilibrium. We consider a halo to be dynamically
``relaxed'' if it simultaneously satisfies all three of the following
conditions \citep{Neto2007}: (i) $f_{\rm sub}<0.1$, (ii) $d_{\rm
  off}<0.07$ and (iii) $2T/|U|<1.35$. Here $f_{\rm sub}$ is the
fraction of the halo's virial mass contributed by subhalos, $d_{\rm
  off}=|\mathbf{r}_p-\mathbf{r}_{\rm CM}|/r_{200}$ is the distance
between the halo barycenter and the location of its potential minimum,
expressed in units of $r_{200}$; and $2T/|U|$ is the virial ratio of
kinetic to potential energies, measured in the halo rest frame. None
of our conclusions are heavily affected by these
restrictions. Unrelaxed systems make up only $20\%$ of all halos with
virial mass of order $10^{12}\, M_\odot$ and $25\%$ of $\sim 10^{13}\,
M_\odot$ halos. Only at very large halo masses, such as cluster-sized
$\sim 10^{14}\, M_\odot$ systems, the unrelaxed fraction exceeds
$50\%$. We refer the reader to \citet{Neto2007} for further
discussion of these criteria, and to \citet{Ludlow2012} for a
discussion of how the inclusion of out-of-equilibrium systems may
impact the mass-concentration relation at large halo masses.

\section{Analysis}
\label{SecAnal}

\subsection{Fitting Formulae}
\label{SecFitForm}

We consider two different formulae to fit halo density profiles. The
NFW profile is given by
\begin{equation}
\frac{\rho(r)}{\rho_{\rm crit}}=\frac{\delta_c}{(r/r_s)(1+r/r_s)^2},
\label{EqNFW}
\end{equation}
where $r_s$ is a scale radius, $\rho_{\rm crit} \equiv 3H^2/8\pi G$ is
the critical density, and $\delta_c$ is the
halo dimensionless characteristic density. These two parameters can also be
expressed in terms of the halo virial mass, $M_{200}$,
and a concentration parameter, $c=r_{200}/r_s$, which is related to $\delta_c$ by
\begin{equation}
\delta_c={200 \over 3} {c^3 \over [\ln(1+c)-c/(1+c)]}.
\label{EqcDeltac}
\end{equation}
Note that for given mass the NFW profile has a single free parameter,
the concentration. This profile can also be expressed in terms of the
enclosed mean density, $M(\langle \rho
\rangle)$, where
\begin{equation}
\langle\rho\rangle(r)={M(<r) \over (4\pi/3) \, r^3} =
{200 \over x^3} {Y(cx) \over Y(c)} \, \rho_{\rm crit}\,,
\label{EqMrhoNFW}
\end{equation}
where $x=r/r_{200}$ and $Y(u)=\ln(1+u)-u/(1+u)$.

The Einasto profile \citep[]{Einasto1965} has an extra free parameter,
the shape parameter $\alpha$, and may be written as
\begin{equation}
\ln\biggl(\frac{\rho_E}{\rho_{-2}}\biggr)=-\frac{2}{\alpha}\biggl[\biggl(\frac{r}{r_{-2}}\biggr)^{\alpha}-1\biggr].
\label{EqEinasto}
\end{equation}
The parameter $r_{-2}$ marks the radius where the logarithmic slope of
the density profile is equal to $-2$. The same property holds for the
NFW scale radius, $r_s$, and therefore, for short, we shall hereafter
refer to the scale radius of either profile as $r_{-2}$. Quantities
measured at (or within) $r_{-2}$ will be denoted by a ``$-2$''
subscript; e.g., $\langle \rho_{-2} \rangle=\langle
\rho\rangle(r_{-2})$. Of course, like NFW, the Einasto profile may
also be expressed in terms of its enclosed mean density profile,
$M(\langle \rho_E \rangle)$.

We note that, for given concentration, an Einasto
profile with $\alpha \approx 0.18$ resembles closely an NFW profile
over roughly two decades in radius or enclosed mass. Profiles with
other values of $\alpha$ deviate systematically from the NFW shape
\citep[see, e.g.,][]{Navarro2004,Navarro2010}.

\begin{figure*}
\begin{center}
\resizebox{13cm}{!}{\includegraphics{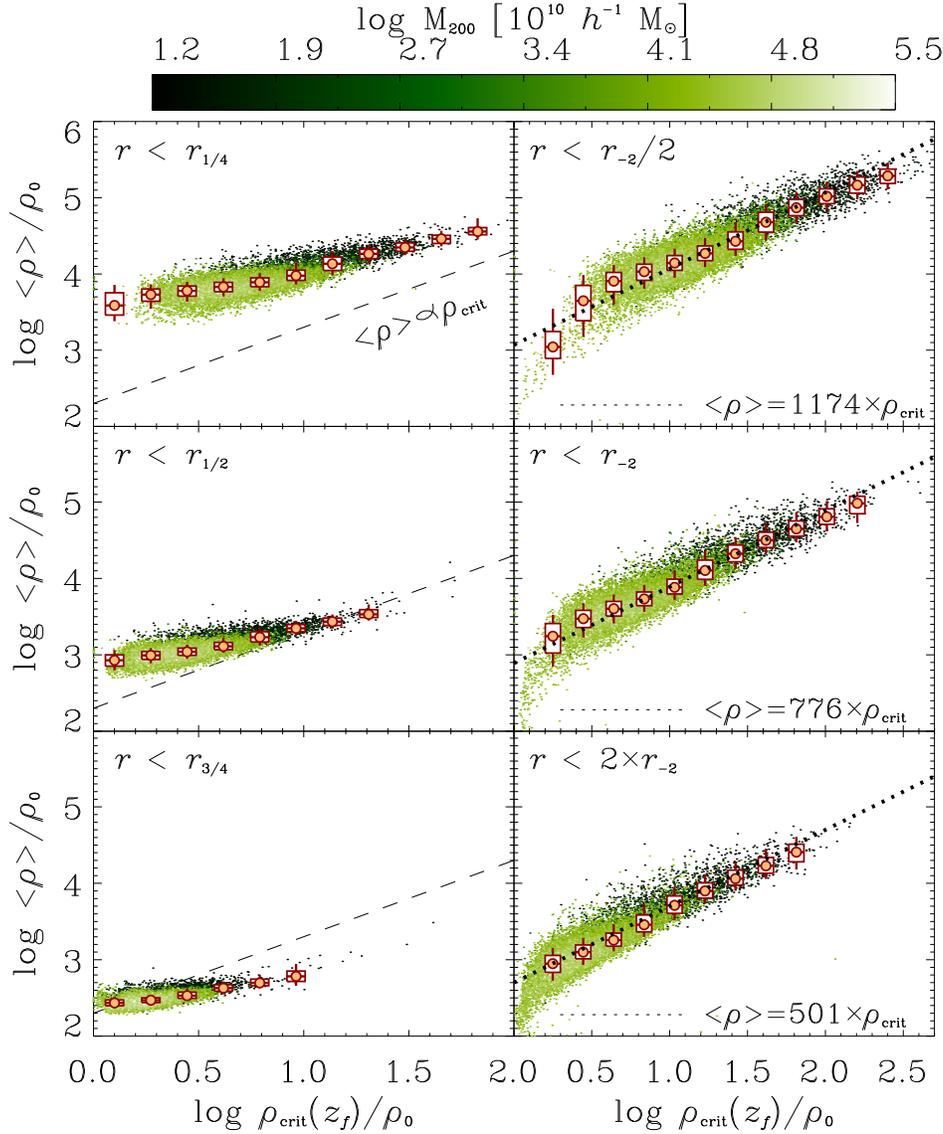}}
\end{center}
\caption{Relation between mass profiles at $z=0$ and accretion
  histories for relaxed, well-resolved halos ($N_{200} >
  2.5\times10^4$) in our sample. Individual halos are colored by mass,
  according to the color bar at the top of the plot. {\it Left panels:} Mean enclosed
  densities within the radii, $r_{1/4}$, $r_{1/2}$, and $r_{3/4}$,
  containing, respectively, $25\%$, $50\%$ and $75\%$ of the virial
  mass, shown as a function of the (critical) density of the Universe
  at the time when the progenitor's virial mass equals the mass
  enclosed within each of those radii at $z=0$. These densities are
  correlated, as expected if denser halos collapse earlier. However,
  the dependence varies with radius and is generally quite weak. This
  explains, for example, why measures of halo density (such as the
  concentration) correlate only poorly with the half-mass formation
  time. Medians, quartiles, and 10/90 percentiles are indicated by the
  box-and-whisker symbols.  {\it Right panels:} 
  As the left panels, but for radii equal to half, one, and
  two times the scale radius, $r_{-2}$. The dotted line indicates
  direct proportionality, scaled vertically to best fit the data of
  each panel (fit parameters given in the legends). The excellent
  agreement between this simple scaling and the data implies that,
  expressed in units of the scale radius, the shape of the mass
  profile of a halo is intimately related to that of the accretion history of
  its main progenitor.}
\label{FigRhoRho}
\end{figure*}

\subsection{Profile Fitting}
\label{SecProfs}

Our analysis deals primarily with the spherically averaged density
profiles of relaxed CDM halos identified at $z=0$ in each MS. We
construct radial profiles using $32$ concentric bins, equally
spaced in $\log r$, spanning the radial range $-2.5\leq \log r/r_{200}
\leq 0$.

The Einasto profile has three free parameters: $\rho_{-2}$, $r_{-2}$,
and $\alpha$. These are simultaneously adjusted in order to minimize
its rms deviation from the binned density profiles. In practice, we
define a figure of merit,
\begin{equation}
  \psi^2 = \frac{1}{N_{\rm bin}} \sum_{i=1}^{\rm Nbin} [\ln \rho_i - \ln \rho_E(\rho_{-2};r_{-2};\alpha)]^2,
  \label{eq:FoM}
\end{equation}
which is minimized to obtain the best-fitting set of parameters for
any given halo. Equation~\ref{eq:FoM} weights equally all
    logarithmic radial bins and, for a given radial range, is
    approximately independent of the number of bins used. It measures
    deviations of the true profile from the model caused by systematic
    shape differences as well as by transient features induced by, for
    example, substructures or tidal streams.  These features lead to
    highly correlated bin-to-bin deviations that typically dominate
    over the Poisson noise in the individual radial bins. For this
    reason we have decided to weight all bins equally \citep[see][for
    further discussion]{Navarro2010}.

In practice, the parameters $\rho_{-2}$ and $r_{-2}$
can be expressed in a variety of equivalent forms, such as virial mass
and concentration ($M_{200}$,$c$), or the magnitude and location of
the peak in the circular velocity curve ($V_{\rm max},r_{\rm
  max}$). In order to ease comparisons with previous work, we
characterize the dark matter halo mass profile in terms of its virial
mass $M_0=M_{200}(z=0)$, its concentration $c=r_{200}/r_{-2}$, and its
Einasto ``shape'' parameter, $\alpha$. The Einasto profile provides an
excellent description of the density profile of relaxed MS halos: the
median value of $\psi$ is just $0.073^{+0.014}_{-0.011}$,
where the range represents the 25th and 75th percentiles.

An analogous procedure is used when NFW fits need to be performed; in
this case, the two parameters estimated by the fit can also be
expressed as the virial mass and concentration.

The fits are carried out over a radial range $r_{\rm min} < r <
r_{200}$. The fitting procedure yields robust estimates for
$\rho_{-2}$, $r_{-2}$ and $\alpha$, provided $r_{\rm min}$ is chosen
to be the minimum of either $r_{\rm conv}$ or $0.05 \times r_{200}$.
Here, $r_{\rm conv}$ is the convergence radius defined by
\citet{Power2003}, where circular velocity profiles converge to better
than $\sim$10\%.

\subsection{Mass Profiles and Accretion Histories}

The left panel of Fig.~\ref{FigRhoProf} illustrates the role of $c$
and $\alpha$ in describing the density profile. This figure shows the
density profile of MS-II halos selected in a narrow range of mass,
$1.24<\log M_{200}/10^{10} h^{-1} M_\odot < 1.54$. (Densities are
weighted by $r^2$ in order to enhance the dynamic range of the plot.)
Each profile corresponds to different systems, grouped by
concentration: green squares track the median\footnote{Median profiles
  are computed at each radius after scaling all individual profiles as
  in Fig.~\ref{FigRhoProf}.} profile of halos with average
concentration for that mass; blue circles and red triangles correspond
to halos with concentration $\sim 50\%$ higher and lower than the
average, respectively (see boxes in the top panel of
Fig.~\ref{FigMcA}).

\begin{figure*}
\begin{center}
\resizebox{16cm}{!}{\includegraphics{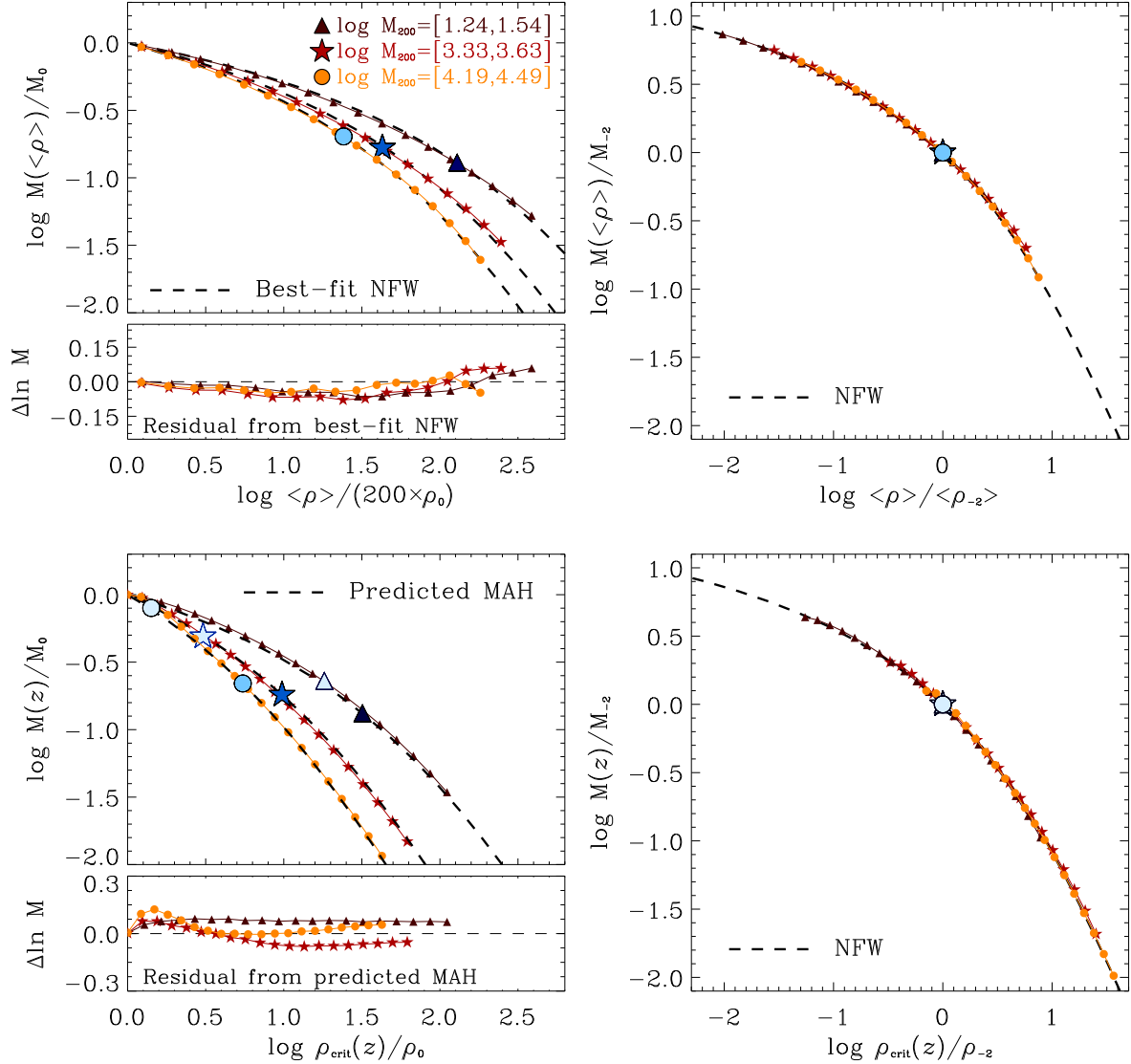}}
\end{center}
\caption{Average mass profiles at $z=0$ and accretion histories for
  halos in three different mass bins (see shaded regions in the bottom
  panel of Fig.~\ref{FigMcA}). {\it Top left:} Average mass profiles
  of all halos in each bin, plotted as enclosed mass (in units of
  $M_{200}$), versus inner density (in units of $200\times$ the
  critical density). Dashed lines are best-fit NFW profiles, which
  have a single adjustable parameter, the concentration, $c=r_{200}/r_{-2}$. 
  Heavy filled symbols indicate the enclosed mass,
  $M_{-2}$, and density, $\langle \rho_{-2} \rangle$, at the scale
  radius of each profile. Residuals from the best fits are shown in
  the bottom inset. {\it Top right:} Same as top-left panel, but
  scaled to the enclosed mass, $M_{-2}$, and overdensity,
  $\langle\rho_{-2}\rangle$, at the scale radius. Scaled in this
  manner, halo mass profiles all look alike and are very well
  approximated by an NFW profile (dashed curve). {\it Bottom left:}
  Average accretion histories of the same halos shown in the top
  panels. The plots show the growth of the virial mass of the main
  progenitor, normalized to the final mass at $z=0$, as a function of
  time, expressed in terms of the critical density of the Universe at
  each redshift. The dashed curves are {\it not} fits to the
  data. Rather, they indicate accretion histories parameterized, as in
  the top panel, by an NFW profile in this $M$-$\rho$ plane. The
  single adjustable parameter to these profiles is fully specified by
  the filled heavy symbols, which indicate $M_{-2}$, chosen to match
  that of the mass profiles (top-left panel) and by $\rho_{\rm
    crit}(z_{-2})$, computed as $776\, \langle \rho_{-2} \rangle$
  following the correlation shown in the middle panel of
  Fig.~\ref{FigRhoRho}. The light-colored heavy symbols 
  indicate the scale mass and density of the predicted NFW profile;
  dark filled symbols mark the location of the halo characteristic 
      mass and the corresponding formation time. {\it Bottom right:} Same
  accretion histories as in the bottom-left panel, but scaled to
  the characteristic values of the MAH: $M_{-2}$ and 
  $\langle \rho_{-2} \rangle$ (light heavy symbols in the
  bottom-left panel). Note the remarkable similarity in the shape of
  the halo mass profiles at $z=0$ and that of the accretion histories
  of their main progenitors.}
\label{FigRhoMaccNFW}
\end{figure*}

In the scaled units of Fig.~\ref{FigRhoProf} the scale radius,
$r_{-2}$, signals the location of the maximum of each curve, and different
concentrations show as shifts in the position of the maxima, which are
indicated by large filled circles. In addition to their different
concentrations, the profiles differ as well in $\alpha$, which
increases with decreasing concentration (see legends in
Fig.~\ref{FigRhoProf}). Arrows indicate the half-mass radius of each
profile. Dot-dashed curves show NFW profiles (whose shape is fixed in
this plot) with the same concentration as the best Einasto fit (solid
lines). The density profile curves more gently than NFW for $\alpha \simlt
0.18$ and less gradually than NFW for $\alpha \simgt 0.18$, respectively.

The (median) mass accretion histories corresponding to the same sets
of halos are shown in the right-hand panel of
Fig.~\ref{FigRhoProf}. We define the mass accretion history (MAH) of a
halo as the evolution of the virial mass of the main progenitor\footnote{The
  main progenitor of a given dark matter halo is found by tracing backwards
  in time the most massive halo along the main branch of its merger tree.},
usually expressed as a function of the scale factor $a=1/(1+z)$, and
normalized to the present-day value, $M_0=M_{200}(z=0)$. As expected,
more concentrated halos accrete a larger fraction of their final mass
earlier on. Filled stars indicate the ``half-mass formation
redshift'', $z_{1/2}$, whereas filled circles indicate $z_{-2}$, the
redshift when the mass of the main progenitor first reaches $M_{-2}$,
the mass enclosed within $r_{-2}$ at $z=0$.

\section{Results}
\label{SecRes}

\subsection{The mass-concentration-shape relations}
\label{SecM200conc}

The top panel of Fig.~\ref{FigMcA} shows the mass-concentration
relation for our sample of relaxed halos at $z=0$. Concentrations are
estimated from Einasto fits, and are color coded by the 
shape parameter, $\alpha$, as indicated by the color bar. Open symbols
track the median concentrations as a function of mass. Thin solid
lines trace the 25th and 75th percentiles of the scatter at fixed mass. Different symbols are used
for the different MS runs, as specified in the legend. Note the
excellent agreement in the overlapping mass range of each simulation,
which indicates that our fitting procedure is robust to the effects of
numerical resolution.

The bottom panel of Fig.~\ref{FigMcA} shows the mass-$\alpha$
relation, colored this time by concentration. The trend is again
consistent with earlier work; the median values of $\alpha$ are fairly
insensitive to halo mass, except at the highest masses, where it
increases slightly. The mass-concentration-shape trends are consistent
with earlier work; for example, the dashed lines correspond to the fitting
formulae proposed by \citet{Gao2008} and reproduce the overall trends
very well.

Fig.~\ref{FigMcA} illustrates an interesting point already hinted at
in Figure~\ref{FigRhoProf}: the shape parameter seems to correlate
with concentration at given mass.  Interestingly, {\it halos of
  average concentration have approximately the same shape parameter}
($\alpha\approx 0.18$, i.e., quite similar to NFW), regardless of
mass. Halos with higher-than-average concentration have smaller values
of $\alpha$ and vice versa. This suggests that the same mechanism
responsible, at given mass, for deviations in concentration from the
mean might also be behind the different mass profile shapes at $z=0$
parameterized by $\alpha$. We explore this possibility next.

\subsection{Characteristic densities and assembly times}
\label{SecRhoRho}

As pointed out by \citet{Navarro1997} and confirmed by
subsequent work \citep[see, e.g.,][]{Jing2000}, the scatter in
concentration is closely related to the accretion history of a halo:
the earlier (later) a halo is assembled the higher (lower) its
concentration. 

This is clear from the assembly histories shown in
Fig.~\ref{FigRhoProf}, which illustrate as well that defining
``formation time'' in a way that correlates strongly and unequivocally
with concentration is not straightforward. For example, the often-used
half-mass formation redshift, $z_{1/2}$, varies only weakly with
$c$, making it an unreliable proxy for concentration
\citep{Neto2007}. {\it An ideal definition of formation time would
  result in a natural correspondence between the characteristic
  density of a halo at $z=0$ and the density of the Universe at the
  time of its assembly.}

We explore two possibilities in Fig~\ref{FigRhoRho}. Here we show 
the mean density enclosed within various characteristic radii at $z=0$
versus the critical density of the Universe at the time when the main
progenitor mass equals the mass enclosed within the same radii.

\begin{figure}
\begin{center}
\resizebox{8cm}{!}{\includegraphics{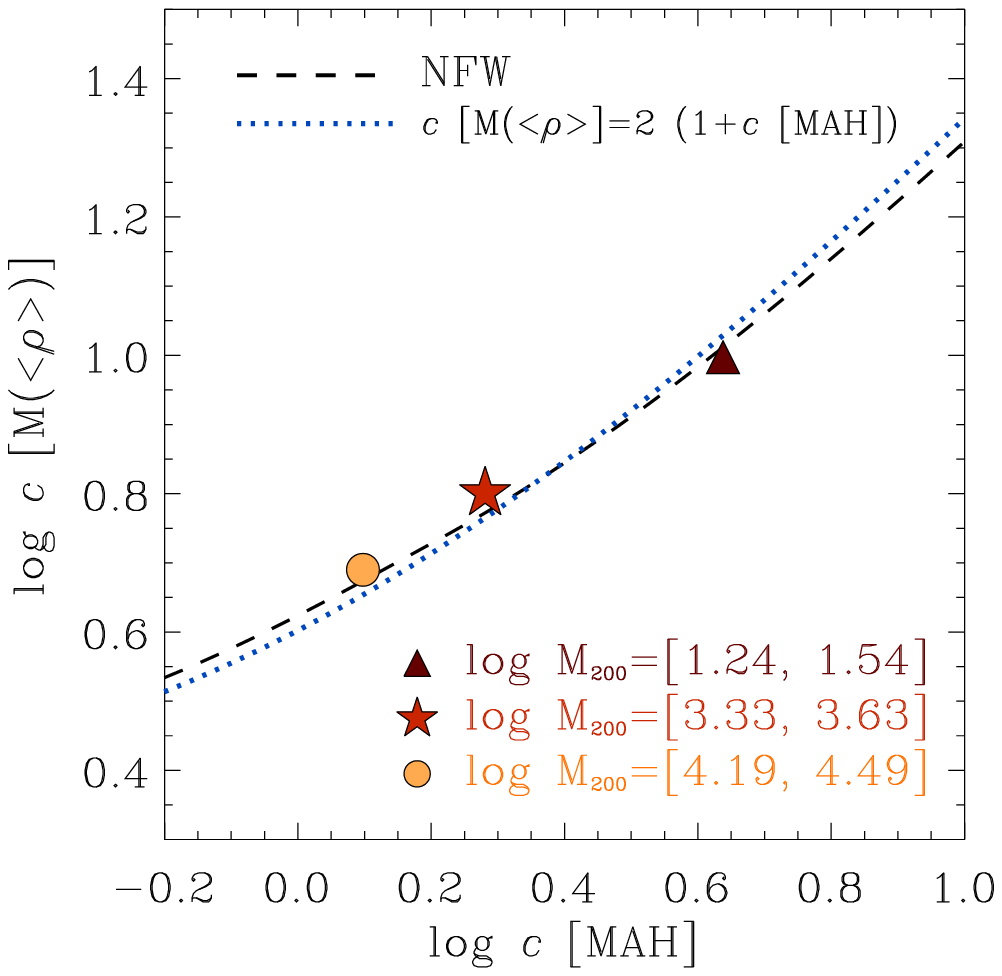}}
\end{center}
\caption{Relation between concentration parameters obtained from NFW
  fits to the average accretion histories and mass profiles shown in
  Fig.~\ref{FigRhoMaccNFW}. The dashed curve indicates the expected
  concentration-concentration dependence given the correlations shown
  in the middle-right panel of Fig.~\ref{FigRhoRho}, assuming an NFW 
  profile. The dotted line shows the best fit obtained using
      Eq.~\ref{fit_cc}; the parameters of the fit are provided in
      Table~\ref{Table1}. Note that the 
  relation is rather shallow, indicating that even halos whose accretion 
  histories differ greatly may have similar concentrations, a result 
  consistent with the weak mass-concentration dependence reported in 
  earlier work.}
\label{FigCC}
\end{figure}

\begin{figure}
\begin{center}
\resizebox{8cm}{!}{\includegraphics{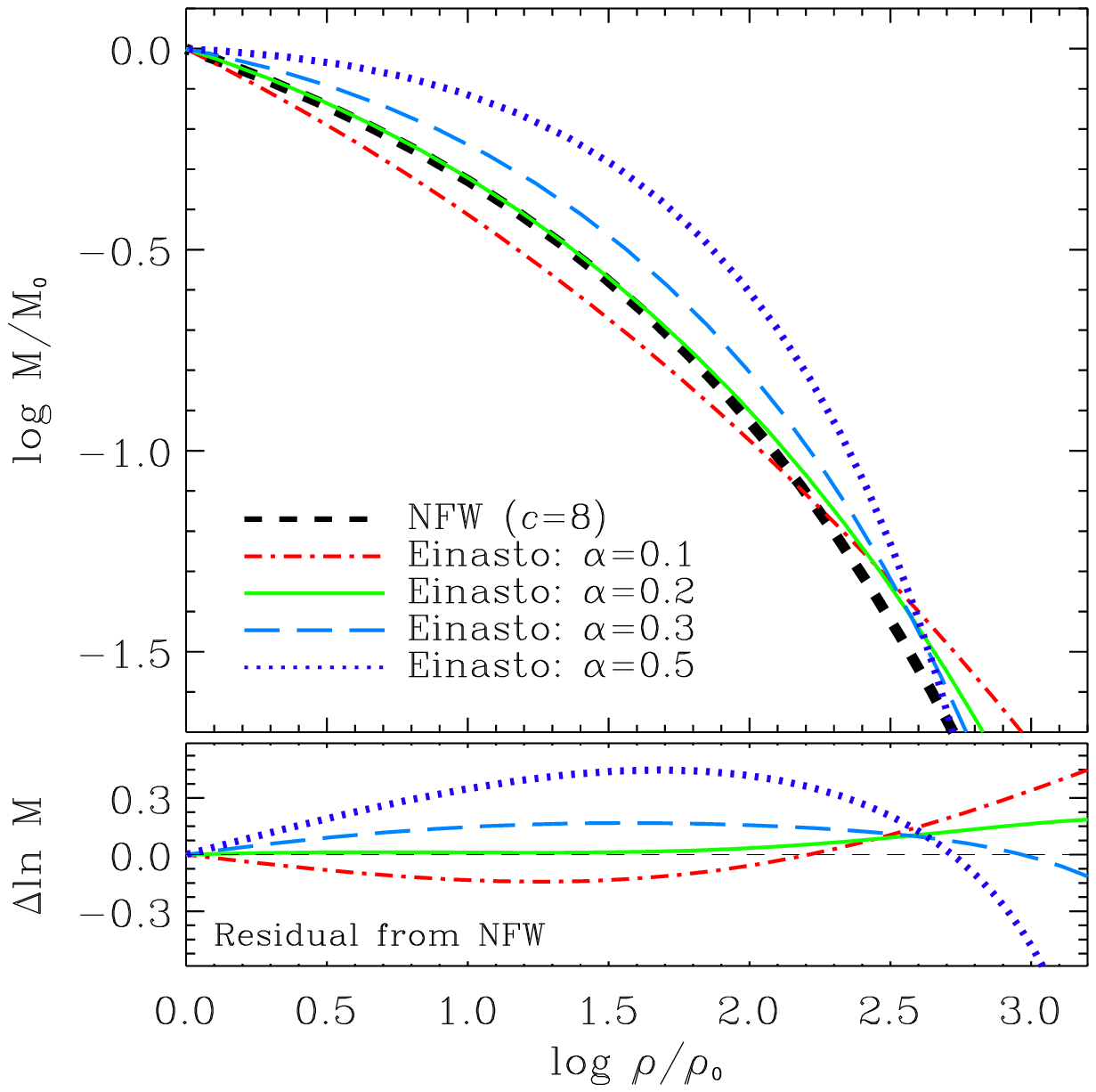}}
\end{center}
\caption{Mass accretion histories, $M(\rho_{\rm crit})$, corresponding
  to Einasto profiles, compared with NFW. Note that NFW resembles
  closely an Einasto profile with $\alpha \sim 0.18$ or so. Larger or smaller
  values of $\alpha$ correspond to halos that have been assembled more
  or less rapidly than the NFW curve, respectively. Residuals from NFW
  are shown in the bottom panel.}
\label{FigMAHEinNFW}
\end{figure}

The left panels correspond to radii enclosing $1/4$, $1/2$, and $3/4$
of the virial mass of the halo. Dots indicate individual halos colored
by halo mass, as shown in the color bar at the top. Boxes and whiskers trace the
10th, 25th, 75th, and 90th percentiles in bins of $\rho_{\rm
  crit}$. Note the tight but rather weak (and non-linear) correlation
between densities at these radii. This confirms our earlier statement
that ``half-mass'' formation times are unreliable indicators of halo
characteristic density: halos with very different $z_{1/2}$ may
nevertheless have similar concentrations.

The right-hand panels of Fig.~\ref{FigRhoRho} show the same density
correlations, but measured at various multiples of $r_{-2}$, the scale
radius of the mass profile at $z=0$. The middle panel shows that the
mean density within $r_{-2}$, $\langle \rho_{-2}
\rangle=M_{-2}/(4\pi/3)r_{-2}^3$, is {\it directly proportional} to
the critical density of the Universe at the time when the virial mass
of the main progenitor equals $M_{-2}$. Intriguingly, this is
also true at $r_{-2}/2$ (top right panel) and at $2\times r_{-2}$
(bottom right panel), although with different proportionality
constants (listed in the figure legends).

This means that there is an intimate relation between the mass profile
of a halo and the shape of its mass accretion history, in the sense
that, once the scale radius is specified, the MAH can be reconstructed
from the mass profile, and vice versa. Since mass profiles are nearly
self-similar when scaled to $r_{-2}$, this implies that accretion
histories must also be approximately self-similar when scaled
appropriately. The MAH self-similarity has been previously discussed by
\citet{vandenBosch2002}, but its relation to the shape of the mass
profile, as highlighted here, has so far not been recognized.

\begin{figure*}
\begin{center}
\resizebox{16cm}{!}{\includegraphics{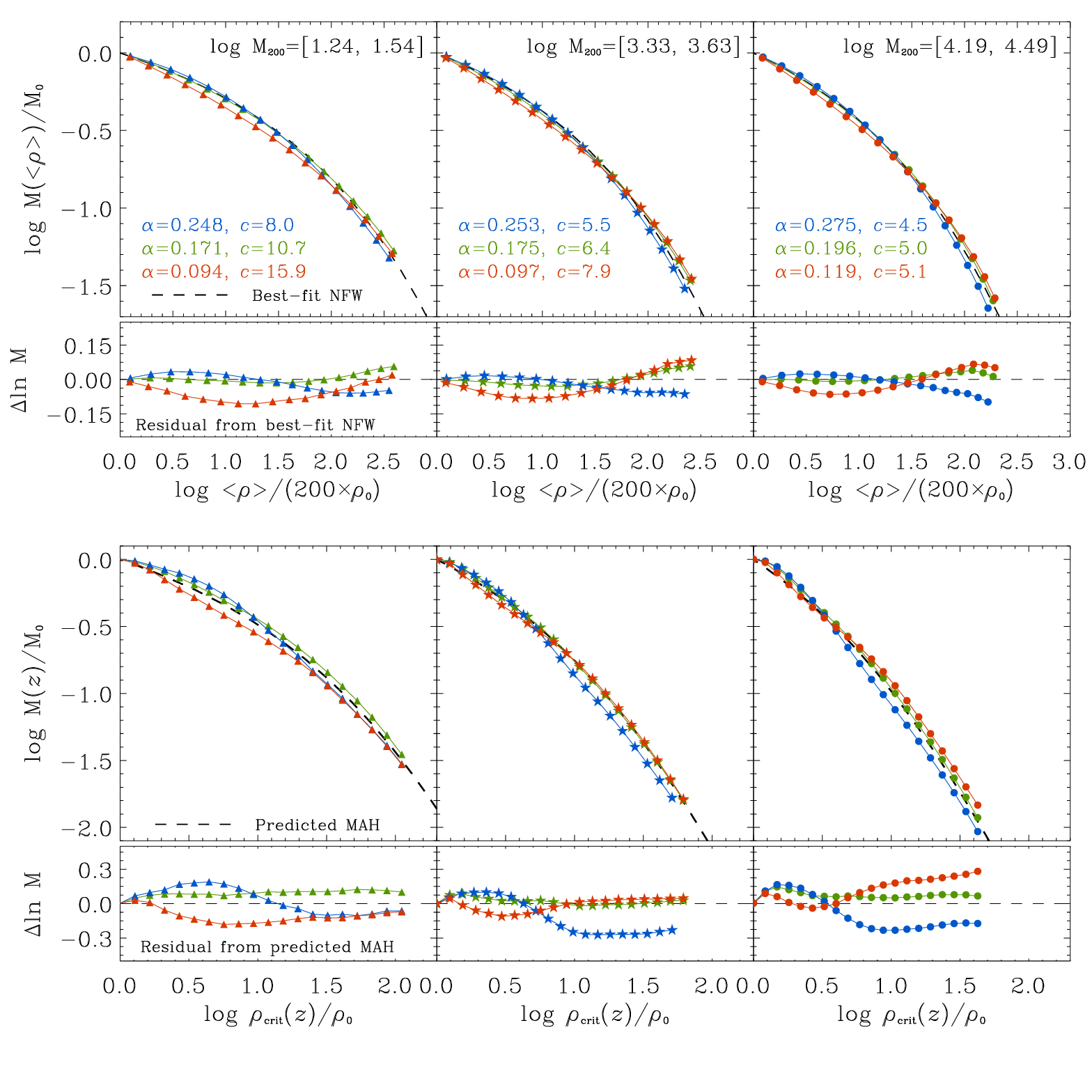}}
\end{center}
\caption{As Fig.~\ref{FigRhoMaccNFW} but for halos with
  higher-than-average (blue), average (green), or lower-than-average
  (red) values of the Einasto parameter $\alpha$ (see boxes in the
  bottom panel of Fig.~\ref{FigMcA}). Left, middle, and right panels
  correspond to each of the three mass bins, as indicated in the
  legends. {\it Top panels:} Average mass profiles compared with the
  best-fit NFW profile for all halos of the same mass (see top left
  panel of Fig.~\ref{FigRhoMaccNFW}). Residuals from that profile are
  shown at the bottom of each panel.  Note the similarity between the
  residual curves of similar color at all masses. Different values of
  $\alpha$ imply different profile shapes, and deviate systematically
  from NFW. {\it Bottom panels:} Average mass accretion histories
  corresponding to the same halos as in the top panels. The dashed
  curves indicate the average ``NFW accretion histories'' for each
  mass bin, as shown in the bottom-left panel of
  Fig.~\ref{FigRhoMaccNFW}. Residuals from this average history are
  shown in the bottom inset of each panel. Note the similarity between
  the shape of the residual curves of similar colors in all
  panels. This indicates that the mass accretion history is intimately
  linked to the mass profile at $z=0$. Halos that, at $z=0$, have mass
  profiles that deviate from NFW in a particular way have accretion
  histories that deviate from the NFW shape in a similar way.}
\label{FigRhoMaccE}
\end{figure*}

\begin{figure*}
\begin{center}
\resizebox{16cm}{!}{\includegraphics{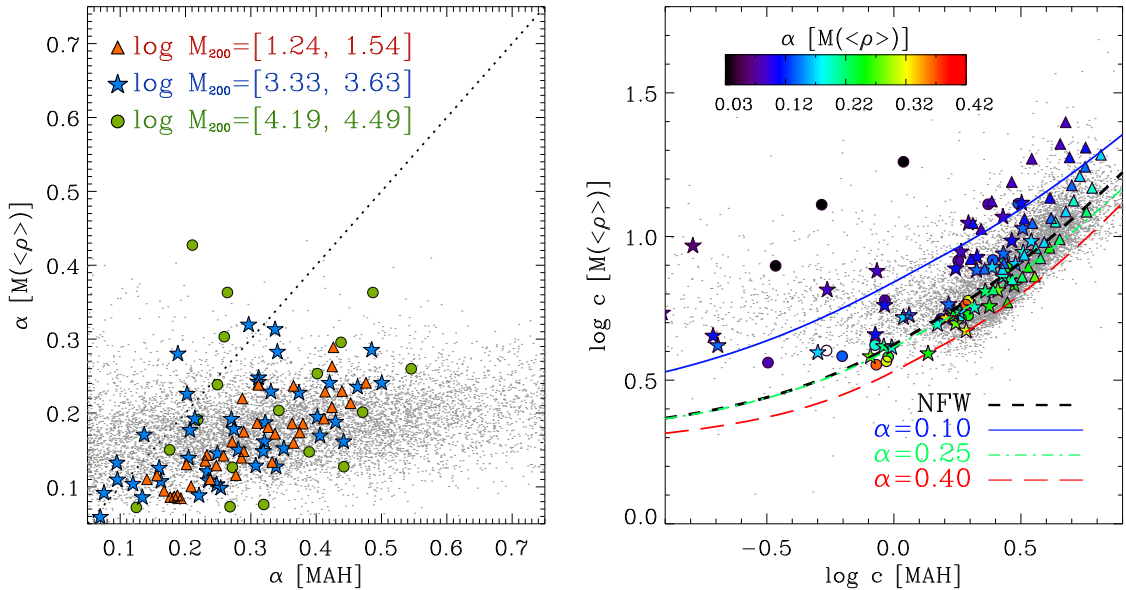}}
\end{center}
\caption{Concentrations and shape parameters of Einasto profiles
  fitted to either accretion histories or mass profiles at
  $z=0$. Heavy symbols correspond to well-resolved halos grouped
  according to the $c$ and $\alpha$ parameters of their mass profile
  (see details in the text). Grey dots correspond to individual halos
  in the same three mass bins chosen fin Fig.~\ref{FigRhoMaccE}.  The
  left panel shows that the shape of the mass accretion history and
  that of the mass profile are correlated. The panel on the right is
  analogous to Fig.~\ref{FigCC} and shows that the same applies to the
  concentrations. In this case, the relation depends on the value of
  $\alpha$, as shown by the colored lines labelled in the legend. The
  heavy symbols are of the same type as in the left panel, but colored
  by $\alpha$ (see inset). Note that the correlations are relatively shallow,
  implying that even large departures from NFW-like mass accretion
  histories lead only to minor deviations from NFW in the mass
  profiles.}
\label{FigccAA}
\end{figure*}

\subsection{NFW accretion histories and mass profiles}
\label{SecNFWMAH}

We explore further the relation between MAH and mass profile by
casting both in a way that simplifies their comparison, i.e., in terms
of mass versus density. In the case of the mass profile, this is just
the enclosed mass-mean inner density relation, $M(\langle \rho
\rangle)$ (see Sec.~\ref{SecFitForm}). For the MAH, this reduces to
expressing the virial mass of the main progenitor in terms of the
critical density, rather than the redshift, $M(\rho_{\rm
  crit}(z))$. In what follows, we shall scale all masses to the virial mass of the halo
at $z=0$, $M_0$; $\rho_{\rm crit}(z)$ to the value at present,
$\rho_0$; and $\langle \rho \rangle$ to $200\, \rho_0$.

The top-left panel of Fig.~\ref{FigRhoMaccNFW} shows, in these scaled
units, the average $M(\langle \rho \rangle)$ profile for halos in
three different narrow mass bins (indicated by grey vertical bars in
the bottom panel of Fig.~\ref{FigMcA}). These mean profiles are
computed by averaging halo masses, for given $\langle \rho \rangle$,
after scaling all individual halos as indicated above. As expected,
each profile is well fit by an NFW profile where the concentration
increases gradually with decreasing mass. The heavy symbols on each
profile indicate the value of $M_{-2}$ and $\langle \rho_{-2}
\rangle$. The top-right panel shows the same data, but scaled to these
characteristic masses and densities. Clearly the three profiles follow
closely the same NFW shape, which is fixed in these units.

The corresponding MAHs, computed as above by averaging accretion
histories of scaled individual halos, are shown in the bottom-left panel of
Fig.~\ref{FigRhoMaccNFW}. The heavy symbols on each profile again
indicate the value of $M_{-2}$ (as in the above panel), as well as
$\rho_{\rm crit}(z_{-2})=776\, \langle \rho_{-2} \rangle$, computed
using the relation shown in the middle-right panel of
Fig.~\ref{FigRhoRho}. 

In these scaled units, a single point can be used to specify the
``concentration'' of an NFW profile, which is shown by the dashed
curves. Interestingly, these provide excellent descriptions of the MAHs:
rescaled to their own characteristic density and mass they all look
alike and also follow closely the NFW shape (bottom-right panel of
Fig.~\ref{FigRhoMaccNFW}). {\it The mass accretion histories and mass
  profiles of CDM halos are not only nearly self-similar: they both
  have similar shapes that may be approximated very well by
  the NFW profile.}

This implies that the concentration of the mass profile just reflects
the ``concentration'' of the MAH. Indeed, assuming that the NFW shape
holds for both, the relation $\rho_{\rm crit}(z_{-2})=776\, \langle
\rho_{-2} \rangle$ delineates a unique relation between the two
concentrations, which is shown as a dashed line in
Fig.~\ref{FigCC}. The three symbols in the same figure
correspond to the three average profiles and MAHs shown in
Fig.~\ref{FigRhoMaccNFW} and clearly follow the same relation. 
The dotted line in Fig.~\ref{FigCC} shows the best-fit relation of the form
\begin{equation}
c \ [{\rm{M}}\langle\rho\rangle]=a_1 \ (1+a_2\times c \ {\rm{[MAH]}})^{a_3},
\label{fit_cc}
\end{equation}
which is accurate to better than $3\%$ over the range $-0.5 \le \log \ c \ {\rm{[MAH]}} \le 1.5$. The best-fit parameters are provided in Table~\ref{Table1}.

\begin{center}
\begin{table}
  \caption{Parameters obtained for best-fits of eq.~\ref{fit_cc} to the
    concentration-concentration relations for NFW profiles and for 
    Einasto profiles with several values of the shape parameter
    $\alpha$. For all cases provided, fits are accurate to better than 
    $\simlt$3\% over the range $-0.5 \le \log \ c \ {\rm{[MAH]}} \le 1.5$ }
\begin{tabular}{c c c c c}\hline \hline
&NFW&&& \\      
& $a_1$   & $a_2$ & $a_3$ & \\ 
& 2.521   & 0.729 & 0.988 & \\ \hline
&Einasto&  & & \\
$\alpha$ & $a_1$   & $a_2$ & $a_3$ & \\ 
 0.10    & 4.124   & 0.849 & 0.833 & \\ 
 0.15    & 3.365   & 0.692 & 0.899 & \\ 
 0.20    & 2.946   & 0.614 & 0.953 & \\ 
 0.25    & 2.697   & 0.557 & 1.003 & \\ 
 0.30    & 2.504   & 0.530 & 1.042 & \\ 
 0.35    & 2.322   & 0.528 & 1.068 & \\ 
 0.40    & 2.154   & 0.543 & 1.084 & \\ 
\hline
\end{tabular}
\label{Table1}
\end{table}
\end{center}

Note that this is consistent with earlier claims that halo
concentration is linked to the time when halo growth switches from a
fast- to a slow-accretion phase \citep[][]{Wechsler2002,Zhao2003a}.
In our interpretation, since both the MAH and the mass profile follow
the same NFW shape the scale radius of one tracks that of the other:
the ``curvature'' of the MAH is therefore reflected in that of the
mass profile. Note as well that the relation shown in Fig.~\ref{FigCC}
is rather weak; in other words, even large changes in the MAH map onto
a small range of concentrations in the mass profiles. This is 
at the root of the weak correlation between concentration and virial
mass reported in earlier work \citep[see, e.g.,][]{Neto2007}.

\subsection{Einasto accretion histories and mass profiles}
\label{SecEMAH}

The striking similarity between the shapes of the MAH and mass profile
discussed above suggests an explanation for why halos that are
outliers in the mass-concentration relation tend to have mass profiles
that differ more significantly from NFW (i.e., they have $\alpha$
parameters that differ from $0.18$, see Fig.~\ref{FigMcA}). In this
interpretation, outliers in $M_{200}$-$c$ have MAH shapes that differ
systematically from the mean, NFW-like shape.

In order to test this, we may use the Einasto formula to fit both MAH
and mass profiles.  Fig.~\ref{FigMAHEinNFW} shows Einasto $M$-$\rho$
profiles for various values of $\alpha$, and compares them with an NFW
profile of the same concentration. (This figure uses the same scalings
as Fig.~\ref{FigRhoMaccNFW}.) As stated earlier, over the range of
mass and density plotted here the NFW profile is essentially
indistinguishable from an $\alpha=0.18$ Einasto profile, but
systematic deviations become apparent for other values of $\alpha$.
Interpreting Fig.~\ref{FigMAHEinNFW} as a mass accretion history, we
see that $\alpha > 0.18$ corresponds to halos that are assembled more
rapidly than expected from the NFW shape. The opposite holds for
$\alpha <0.18$. This behaviour is clearly seen in the residuals from
NFW, which are shown in the bottom inset of the figure.

The top panels of Fig.~\ref{FigRhoMaccE} show the average $M(\langle
\rho \rangle)$ profiles of halos in three different narrow mass bins
chosen to have different values of $\alpha$. These are halos whose
Einasto parameters fall in the boxes drawn in the bottom panel of
Fig.~\ref{FigMcA}. The best-fit NFW profile for each mass bin (as in
Fig.~\ref{FigRhoMaccNFW}) is indicated by a dashed curve in each
panel. Deviations from the NFW curve are shown in the residuals
panels.  As expected, the residuals have different shapes
depending on the value of their shape parameter $\alpha$.

The bottom panels of Fig.~\ref{FigRhoMaccE} show the corresponding
average mass accretion histories and compare them with the mean {\it
  predicted} MAHs shown in the bottom panels of
Fig.~\ref{FigRhoMaccNFW}. The latter are the NFW MAHs that result from
the $\langle \rho_{-2} \rangle$-$\rho_{\rm crit}(z_{-2})$ correlation
shown in Fig.~\ref{FigRhoRho}. The residuals from this predicted MAH
are clearly similar in shape to those in the top panels: in other
words, on average, {\it halos whose mass profiles deviate from NFW
  have mass accretion histories that deviate from the NFW shape in a
  similar way}.

Quantitatively, this implies that the best-fit Einasto parameters of
both MAHs and mass profiles must be correlated. Since we expect the
correlations to be weak (see, e.g., Fig.~\ref{FigCC}) we group halos
by mass, concentration, and shape parameter (as measured from their
mass profiles) before fitting Einasto profiles to their corresponding
average mass accretion histories. To prevent possible biases induced
by numerical-resolution effects the grouping is such that we retain
only well-resolved halos with similar numbers of particles,
$25,000<N_{200}<50,000$. Statistical fluctuations are reduced by
averaging over groups of at least $25$ halos, using a grid in the
$c$-$\alpha$ plane with a mesh of width $\delta \log c = 0.079$ and
$\delta \alpha=0.026$.

The results are shown by the heavy symbols in Fig.~\ref{FigccAA}. The
left panel shows that the MAH shape parameter is clearly correlated
with the shape parameter of the mass profile. Symbols of different
colors are used for halos in each of the three mass bins, which
correspond to different MS. Note that they all delineate the same
trend, despite the fact that they span a range of roughly four decades
in virial mass. Note as well that parameters corresponding to
individual halos (shown as grey dots in the figure) correlate less
well, and that the scatter is larger for the MS-XXL halos. This is
because individual MAHs are often quite complex, especially when major
mergers are involved or halos are recently assembled and still
unrelaxed (even though they pass the relaxation criteria set out in
Sec.~\ref{SecHaloCats}), as is the case for many MS-XXL systems
\citep{Ludlow2012}. These MAHs cannot be well approximated by the
Einasto shape, thus hindering the interpretation of their fit
parameters. We therefore focus the discussion on the parameters fit to
the averaged profiles, shown with heavy symbols in Fig.~\ref{FigccAA}.

The relation between shape parameters is quite weak (the $1$:$1$
relation is indicated by the dotted line), implying that large
variations in MAH shape map onto a narrower range of mass profile
shapes. As a result, even halos that assemble early and over a very
short period of time, such as those whose MAH is characterized by
$\alpha\sim 0.5$ (see Fig.~\ref{FigMAHEinNFW}), end up with
$\alpha\sim 0.25$ mass profiles that differ only slightly from
NFW. This is consistent with earlier findings that halos assembled
monolithically and without protracted accretion, such as those formed
in hot dark matter universes, are nevertheless well approximated by
NFW profiles \citep[e.g.,][]{Huss1999,Wang2009}.

It also explains why the NFW profile fits rather well halos formed in
hierarchical scenarios other than LCDM. For example, the accretion
histories of halos formed in scale-free scenarios characterized by a
power-law spectrum of density fluctuations, $P(k)\propto k^n$, depends
on $n$, but rather weakly. For given $n$, the MAH shapes are also, on
average, independent of halo mass. Fitting Einasto profiles to
accretion histories taken from \citet{Zhao2009}, we find that halos
that form from white-noise spectra ($n=0$) have MAHs well described by
$\alpha \sim 0.1$. For $n=-2$, on the other hand, the average MAH shape
is roughly $\alpha \sim 0.2$. These different MAHs result in only a subtle change
in mass profile (see left panel of Fig.~\ref{FigccAA}), which would
have been undetectable at the numerical resolution probed by earlier
work. It may, however, be behind the claim by \citet{Knollmann2008}
that the inner slope of the density profile varies systematically with
the spectral index $n$.

The right-hand panel of Fig.~\ref{FigccAA} shows, on the other hand,
the correlation between the best-fit Einasto concentration parameters
of the MAH and mass profiles, for the same set of halos shown in the
left-hand panel of the same figure. This is analogous to
Fig.~\ref{FigCC}, but for Einasto, rather than NFW,
concentrations. (The NFW $c$-$c$ correlation is shown by a dashed
line.) Because of the extra parameter, the relation between Einasto
concentrations depends on $\alpha$, and is indicated by the colored
lines in the figure for three values of $\alpha$. The symbol types in
both panels are the same, but are colored by $\alpha$ in the
right-hand panel (see color bar inset). Note that, for given $\alpha$,
the $c$-$c$ relation for MS halos follows closely the expected
correlations. As for the NFW profile, the concentration-concentration 
    relation for Einasto profiles can also be approximated by Eq.~\ref{fit_cc}. In
    Table~\ref{Table1} we provide the best-fit parameters obtained by fitting
    Eq.~\ref{fit_cc} to these relations for several different values of $\alpha$.

As in the left panel, grey dots in the right-hand panel of
Fig.~\ref{FigccAA} correspond to fits to individual halo MAH and mass
profiles. These clearly follow the same trend as the averaged profiles
but with larger scatter. In particular, complexities in the MAH caused
by major mergers result at times in extreme values for the
``concentration'' measured from accretion histories. These affect in
particular large mass halos that have recently been assembled. As
discussed by \citet{Ludlow2012}, many such halos pass the relaxation
criteria but are actually out of equilibrium and in a particular phase
of their virialization process. These halos are actually responsible
for most of the outlier points in Fig.~\ref{FigccAA}. Although 
a detailed investigation of the accuracy with which our model can predict 
the concentrations of {\em individual} halos from their MAHs is beyond the
scope of this paper, we plan to return to this subject in forthcoming work.

We conclude that there is strong evidence for a link between the
concentration and shape of the mass profile of halos and their
accretion histories. The correlations are well-defined but weak, in
the sense that even MAHs whose shapes deviate substantially from the
mean lead to halos that depart only subtly from the average, NFW-like
mass profile. This is probably due to the virialization process, where
non-linear effects lead to a substantial but incomplete erasure of
memory of the initial conditions from the equilibrium structure of a
halo. The structural similarity of CDM halos thus seems to arise from the
mass-independence of MAH shapes aided by the homogenizing effect of
halo virialization.

\section{Summary and Conclusions}
\label{SecConc}

We have examined the mass profile and accretion histories of
equilibrium cold dark matter halos identified in the Millennium
Simulation series.  As reported in earlier work, halo mass profiles
are well approximated by NFW profiles which, at given virial mass, are
characterized by a single parameter, the concentration,
$c=r_{200}/r_{-2}$. Although in general deviations from the NFW
profile are small, improved fits may be achieved using Einasto
profiles characterized, at given virial mass, by the concentration and
an extra shape parameter, $\alpha$.

Our main finding is that these parameters are strongly linked to the
accretion history of a halo. The mean density within the scale radius,
$r_{-2}$, is directly proportional to the critical density of the
Universe at the time when the main progenitor's mass equals that
within $r_{-2}$. Scaled to these characteristic values of mass and
density the shape of the mass accretion history, expressed as
$M(\rho_{\rm crit}(z))$, is, on average, independent of halo
mass. Furthermore, this shape is nearly identical to that of the
enclosed mass-mean inner density profile ($M(\langle \rho \rangle)$)
of the halo at $z=0$, which can be well approximated by the NFW
profile.

This result suggests that the structural similarity of halos of
different mass is related to the fact that their accretion
histories are independent of mass. It also clarifies how the accretion
history determines the concentration of a halo; since accretion
history and mass profile follow the same NFW shape, there is a unique
correspondence between the ``concentration'' parameters of either one,
as shown in Fig.~\ref{FigCC}.

This conclusion is strengthened by the finding that halos whose mass
profiles deviate from NFW and are better approximated by Einasto
profiles, have mass accretion histories that deviate from the NFW
shape in a similar way. This suggests that the extra shape parameter
of the Einasto profile arises because some halos have accretion
history shapes that differ from NFW. Indeed, fitting Einasto profiles
to both $M(\langle \rho \rangle)$ and $M(\rho_{\rm crit}(z))$ yields
correlated concentration and shape parameters.  The correlations are
clear but weak, implying that only halos whose accretion histories
deviate strongly from the NFW shape would have mass profiles that
deviate noticeably from the average, NFW-like shape. We ascribe this
result to the convergent effects of virialization, which partially
erase the memory of initial conditions from the halo structure.

Our results suggest that the density profiles of halos formed in
hierarchical scenarios other than CDM or monolithically, as in a hot
dark matter Universe, are not truly self-similar. The deviations from
similarity, however, are subtle, and a dedicated program of
high-resolution numerical simulations is needed in order to validate
this prediction.

We may use these findings to predict the dependence of halo
concentration on mass and redshift, as well as the influence of
varying the spectral index or the cosmological parameters, provided
that realistic accretion histories are available, either through
direct numerical simulation or through well-tested semi-analytic modeling
\citep[e.g.,][]{Monaco2002,vandenBosch2002,Zhao2009}. Such studies 
would help to reveal any shortcomings in our
interpretation and should shed further light onto the mechanisms
responsible for CDM halo structure. We plan to address these issues in
a forthcoming paper.

Finally, our findings provide some endorsement to the many previous
studies that have sought a link between the final structure of a halo
and its evolutionary history \citep[e.g.,][]{Bullock2001,Wechsler2002,
Alvarez2003,Zhao2003a,Tasitsiomi2004b,Lu2006,Zhao2009,Wong2012} but still 
fall short of providing a full account of what determines the structure of 
dark matter halos. Hopefully, the link with accretion history we describe
here will help to guide future theoretical work in order to unravel
the mechanism at the root of the remarkable structural similarity of
dark matter halos.

\section*{acknowledgements}
We would like to thank Gerard Lemson for useful discussions, and the Virgo Consortium 
for access to the MS data. We would also like to thank the anonymous referee
for a prompt report that helped improve our paper. ADL acknowledges financial support 
from the SFB (956) from the Deutsche Forschungsgemeinschaft. REA is supported by 
Advanced Grant 246797 “GALFORMOD” from the European Research Council. MB-K acknowledges 
support from the Southern California Center for Galaxy Evolution, a multi-campus 
research program funded by the University of California Office of Research. VS 
acknowledges partial support by TR33, 'The Dark Universe', of the Deutsche 
Forschungsgemeinschaft.

\bsp
\label{lastpage}

\bibliographystyle{mn2e}
\bibliography{paper}

\end{document}